\documentclass[aps,twocolumn,floats,nofootinbib,prl]{revtex4}
\usepackage{graphics,graphicx,epsfig}
\usepackage{amssymb,color}
\usepackage{epsf,epstopdf,wrapfig}
\usepackage {amsmath}

\newcommand{\beq}{\begin{equation}}
\newcommand{\eeq}{\end{equation}}
\newcommand{\beqn}{\begin{eqnarray}}
\newcommand{\eeqn}{\end{eqnarray}}

\begin{document}

\title{Trading bits in the readout from a genetic network}

\author{Marianne Bauer,$^{1-4}$  Mariela D.~Petkova,$^5$ Thomas Gregor,$^{1,2,6}$ Eric F.~Wieschaus,$^{2-4}$   and William Bialek$^{1,2,7}$}

\affiliation{$^1$Joseph Henry Laboratories of Physics, 
$^2$Lewis--Sigler Institute for Integrative Genomics, 
$^3$Department of Molecular Biology, and 
$^4$Howard Hughes Medical Institute, Princeton University, Princeton, NJ 08544 USA\\
$^5$Program in Biophysics, Harvard University, Cambridge, MA 02138, USA\\
$^6$Department of Developmental and Stem Cell Biology, UMR3738, Institut Pasteur, 75015 Paris, France\\
$^7$Initiative for the Theoretical Sciences, The Graduate Center, City University of New York, 365 Fifth Ave, New York, NY 10016}

\begin{abstract}
In genetic networks, information of relevance to the organism is represented by the concentrations of transcription factor molecules.  In order to extract this information the cell must effectively ``measure'' these concentrations, but there are physical limits to the precision of these measurements.  We explore this trading between bits of precision in measuring concentration and bits of relevant information that can be extracted, using the gap gene network in the early fly embryo as an example.  We argue that cells in the embryo can extract all the available information about their position, but only if the concentration measurements approach the physical limits to information capacity. These limits necessitate the observed proliferation of enhancer elements with sensitivities to combinations of transcription factors, but fine tuning of the parameters of these multiple enhancers is not required.
\end{abstract}

\maketitle


Cells control the concentrations of proteins in part by controlling the transcription of messenger RNA.  This control is effected by the binding of transcription factor (TF) proteins to specific sites along the genome.  Transcription factors thus can regulate the synthesis of other TFs, forming a genetic network.  The dynamical variables at  the nodes of this network are the TF concentrations, which represent information of relevance to the cell.  What must the cell do in order to extract and use this information?

We usually think of transcription factors as controlling the level of gene expression, but we can also view the expression level as being the cell's measurement of the TF concentration \cite{bialek+setayeshgar_05,tkacik+al_08a}. As outside observers of the cell, we can measure the concentration of transcription factors with considerable accuracy  \cite{dubuis+al_13b}.
However the cell's ``measurement'' of  TF concentration is based on the arrival of these molecules at their binding sites, and this is a noisy process, 
because TF concentrations are low, in the nanoMolar range  \cite{gregor+al_07b,dostatniBicoid1,drocco2011,hannon2017,KeenanShvartsman2020}.  
Physical limits to the measurement of such low concentrations were first explored in the context of bacterial chemotaxis \cite{berg+purcell_77}, but have proven to be much more general \cite{bialek+setayeshgar_05,kaizu+al_14,friedlander+al_16,mora+nemenman_19}.  What will be important for our discussion is not the precise values of these limits, but rather that the limits exist and are significant on the scale of biological function.   

We focus on the example of the gap genes that are crucial in the early events of embryonic development in fruit flies \cite{nusslein-vollhard+wieschaus_80,jaeger_11}.  These four genes form a network with inputs from primary maternal morphogen molecules, and outputs in the striped patterns of pair-rule gene expression.  These stripes are positioned with an accuracy of $\pm 1\%$ along the long (anterior--posterior) axis of the embryo, and this is the accuracy of subsequent developmental events such as the formation of the cephalic furrow \cite{dubuis+al_13,Liu2013}. The local concentrations of the gap gene proteins provide just enough information to support this level of precision \cite{dubuis+al_13}, and the algorithm that achieves optimal readout of this positional information predicts, quantitatively, the distortions of the pair-rule stripes in mutant flies where individual maternal inputs are deleted \cite{petkova+al_19}.


The gap gene network offers us the chance to ask  how accurately the transcription factor concentrations need to be measured to retain crucial biological information: the information that the gap genes convey about position along the anterior--posterior axis is what allows nuclei  to make distinct cell fate decisions required for development.  We start with more traditional views of how information is represented in TF concentrations, and then argue for a more abstract formulation of the problem.


The classical view of the gap genes is that they are expressed in domains \cite{jaeger_11}.  Implicitly this suggests that fine scale variations in the concentration of these molecules is not important; rather all that matters whether expression is on or off.  The quantitative version of this idea is that subsequent events are sensitive to whether expression levels are above or below a threshold, corresponding to whether a cell is inside or outside an expression domain.  We know  that such simple thresholding omits a lot of the information that gap gene expression levels carry about position along the anterior-posterior axis \cite{dubuis+al_13}.

\begin{figure*}
\centerline{\includegraphics[width = \linewidth]{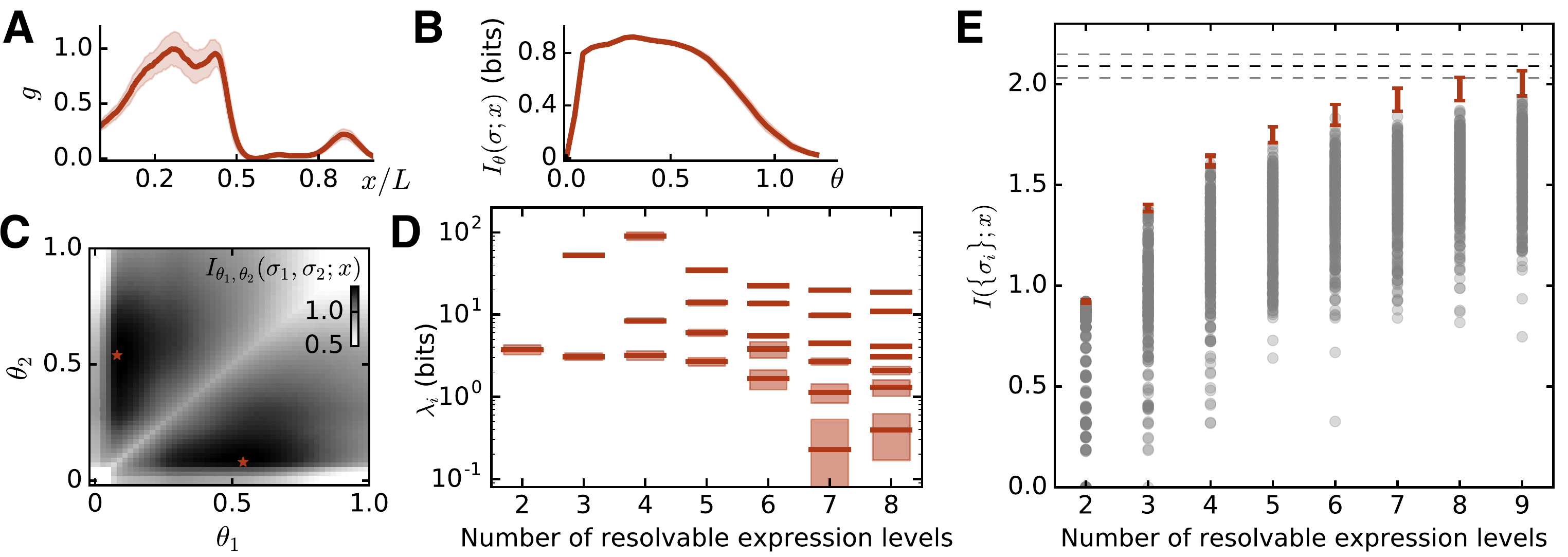}}
\caption{(A) Hb expression level vs position along the anterior--posterior axis embryo. Mean (line) $\pm$ one standard deviation (shading) across $N_{\rm em} = 38$ embryos in a five minute window (40--44 min) in nuclear cycle 14 \cite{petkova+al_19}. 
(B) Positional information vs threshold, from Eq (\ref{Itheta}). (C) Positional information with two thresholds, $I_{\theta_1 \theta_2} (\sigma_1, \sigma_2 ; x)$ (bits).  
	(D) Eigenvalues $\{\lambda_{\rm i}\}$ of the Hessian matrix $\chi$, from Eq (\ref{hessian}). The number of eigenvalues is the number of thresholds, one less than the number of resolvable expression levels.  Shaded bands are $\pm$ one standard error in our estimates. (E) Positional information captured with ${\rm i} = 1,\, \cdots ,\, K$ thresholds, as a function of the number of resolvable levels $K+1$.  Error bars (red) are mean $\pm$ one standard error of our estimate of the maximum. Circles (grey) are 300 values of $I(\{\sigma_{\rm i}\}; x)$ at random settings of the $K$ thresholds $\{\theta_{\rm i}\}$. The black dashed line indicates the positional information $I(g;x)$, available from the expression level if measured precisely, and gray dashed lines are $\pm$ one standard error in our estimate of this information.
\label{Fig01}}
\end{figure*}

In Figure \ref{Fig01}A and B we use  the gap gene {\em hunchback} (Hb) to illustrate this information loss.  At each point $x$ there is an expression level $g$, drawn from a probability distribution $P(g|x)$; looking at many embryos we have samples out of this distribution.  If cells are sensitive only to whether expression levels are above or below a threshold $\theta$, then the variable which  matters is 
\begin{equation}
\sigma = H (g - \theta),
\end{equation}
where $H$ is the Heaviside step function,  $H(y>0)=1$ and $H(y<0)=0$. Then we can estimate the distribution $P_\theta(\sigma | x)$, where the subscript reminds us that the result depends on our choice of threshold,
\begin{eqnarray}
P_\theta(\sigma =1 | x) &=& \int dg\, H (g - \theta) P(g|x)\\
P_\theta(\sigma =0 | x) &=& 1 - P_\theta(\sigma =1 | x) .
\end{eqnarray}
Finally we compute the amount of information that the discrete variable $\sigma$ provides about position,
\begin{equation}
I_\theta(\sigma; x ) = \sum_\sigma\int dx \, P(x) P_\theta(\sigma |x) \log_2\left[ {{P_\theta(\sigma |x)}\over{P_\theta(\sigma)}} \right] \,{\rm bits},
\label{Itheta}
\end{equation}
where a priori all positions along the length of the embryo are equally likely, $P(x) = 1/L$, and 
\begin{equation}
P_\theta(\sigma ) = \int dx \,P(x) P_\theta(\sigma |x) .
\end{equation}
It is important that in exploring the impact of thresholding we allow for the best possible choice of the threshold $\theta$, which in this example proves to be at $\theta^* \sim 1/3$ of the maximal mean expression level.

If the expression level is represented only by the on/off or binary variable $\sigma$, then it can provide at most one bit of information (about anything), and we see that the information about position comes close to this bound, with $I_{\rm max}(\sigma ; x) = 0.92 \pm 0.01\,{\rm bits}$.  But this is less than one half of the information that is carried by the continuous expression level, $I(g;x) = 2.09\pm 0.06\,{\rm bits}$.

One path to recovering the information that was lost by thresholding is to imagine that the cell can resolve more details, perhaps distinguishing reliably among three or four different expression levels rather than just two.  This is equivalent to imagining that we have multiple readout mechanisms, each of which can only distinguish on/off, but the different mechanisms can have different thresholds, in the spirit of the ``French flag'' model \cite{wolpert_69}.  Because we can always put the thresholds in order, having $K$ binary switches is the same as distinguishing $K+1$ different expression levels.  It can be useful to think of thresholding as being implemented at individual binding sites for the TFs, or perhaps at cooperative arrays of binding sites in enhancers, but it is important that our arguments are independent of these microscopic details.

If we have two different  elements, each of which reports on whether the expression level is above or below a threshold, then the relevant variables are
\begin{eqnarray}
\sigma_1 &=& H (g - \theta_1)\\
\sigma_2 &=& H (g - \theta_2).
\end{eqnarray}
We see in Fig \ref{Fig01}C that there is a broad optimum in the positional information that these variables capture, $I_{\theta_1\theta_2}(\{\sigma_1 , \sigma_2 \}; x)$,  when the two thresholds are quite different from one another, $\theta_1^* = 0.1$ and $\theta_2^* = 0.58$; these bracket the optimal single threshold $\theta = 0.34$. The maximum information now is $I_{\rm max}(\{\sigma_1 , \sigma_2 \}; x)  = 1.4\pm 0.015\,{\rm bits}$, noticeably more than in the case with one threshold but still far from capturing all the available information.

We can generalize this idea to multiple thresholding elements, which are described by a set of variables $\{\sigma_{\rm i}\}$, with each $\sigma_{\rm i} = H (g - \theta_{\rm i})$, for ${\rm i} = 1,\, 2,\, \cdots ,\, K$;  the relevant quantity now is $I(\{\sigma_{\rm i}\}; x)$. This positional information depends on all the thresholds $\{\theta_{\rm i}\}$, and we perform a multidimensional optimization   to find the maximum of  $I(\{\sigma_{\rm i}\}; x)$.  Figure \ref{Fig01}E shows that for cells to extract all the positional information available from the Hb concentration, they must distinguish eight or nine different expression levels, in effect representing the expression level with $\sim 3\,{\rm bits}$ of precision.

Distinguishing eight levels in this simple threshold picture requires the cell to set seven thresholds.  It might seem as though this necessitates setting each threshold to its optimal value, a form of fine tuning.  To explore this we choose thresholds at random, uniformly in the relevant interval $0 < \theta < 1$.  As shown in Fig \ref{Fig01}E,  the typical random choices are far below the optimum, as expected.  But  Figures \ref{Fig01}B and C show that there is broad plateau in information vs one or two thresholds, and even with eight thresholds  we find that more than $1$ in $1000$ of our  random choices come within error bars of the optimum.

Another way of looking at the issue of fine tuning is to   examine the behavior of the information in the neighborhood of the optimum,
\begin{equation}
I(\{\theta_{\rm i}\}) = I_{\rm max}(K) + {1\over 2}\sum_{{\rm i,j}=1}^K (\theta_{\rm i} - \theta_{\rm i}^*) \chi_{\rm ij} (\theta_{\rm j} - \theta_{\rm j}^*) + \cdots ,
\label{hessian}
\end{equation}
estimating the Hessian matrix $\chi$ numerically from the data. The matrix $\chi$ has units of bits, as we chose the thresholds to be dimensionless. The eigenvectors of $\chi$ determine the combinations of thresholds that have independent effects on the information, and the eigenvalues $\{\lambda_{\rm i}\}$ of $\chi$ (also in bits) determine the sensitivity along these independent directions.   As the number of thresholds increases we find a wider spread of eigenvalues,  as in a wide class of ``sloppy models'' \cite{gutenkunst+al_07,transtrum+al_15}.  This means that some combinations of thresholds are two orders of magnitude more important than others.

In more detail we find that the eigenvector with the largest eigenvalue is concentrated on the lowest threshold, while higher thresholds are much more loosely constrained.   Although we should again be cautious about overly detailed molecular interpretations, it is natural to think of the mapping $g \rightarrow \{\sigma_{\rm i}\}$ as being implemented by binding of the transcription factor to specific sites along the genome, so that thresholds are set by the binding constants or affinities of the TF for these sites.  The spectrum of the Hessian tells us that the affinity at the strongest binding site must be set carefully, but the weaker binding sites can be scattered more freely across the available dynamic range of concentrations.  A near optimal array of thresholds could thus evolve by duplication of a strong binding binding site, followed by sequence drift to weaker binding, and then  selection for the more complex and reproducible patterns that result from capturing more  positional information, as  in  Ref \cite{francois+siggia_10}.

The idea of the cell taking thresholds is that it is not possible to make arbitrarily fine distinctions, so we imagine that only ``on'' and ``off'' are reliably distinguishable. But thresholding doesn't quite capture our intuition about limited precision because the threshold itself is sharp:  taking a threshold means making perfect distinctions between concentrations that are just above and just below threshold.  We need to find a more precise way of implementing the intuition that cells cannot make arbitrarily precise measurements.

One path to a more realistic view of limited precision is to explore  microscopic models for the interactions between TF proteins and their binding targets along the genome, and for the mechanisms by which binding events influence transcription \cite{bintu+al_05a,segal+al_08,tkacik+al_09,pulkkinen+metzler_13,furlong+levine_18,sabari+al_18}. But then our conclusions depend on our choice of model, and almost certainly our model will be wrong in detail. Instead we can try to abstract away from concrete models, hoping to capture the essence of limited precision in a more general way.

As an example, suppose  that transcription factors influence transcription only when they are bound to their target sites.   In this case the cell does not have direct access to the concentration $g$ but only to the binding site occupancy  averaged over time; there could also be multiple binding sites.  Abstractly, the concentration $g$ is being mapped into some other variable, and we can call this variable $C$,  without worrying about details.  The crucial consequence of noise in the binding events is that $C$ can  provide  only a limited amount of information about the concentration $g$ \cite{tkacik+al_08a}.  Different molecular mechanisms generate different mappings $g \rightarrow C$, but in all mechanisms the low concentrations of the relevant molecules limit the information that is transmitted, $I(C; g)$.

What is relevant for the embryo is information about the position of cells along the anterior--posterior axis.  Thus it is natural to ask: given physical limits on $I(C; g)$, what is the maximum amount of positional information, $I(C;x)$, that can be captured? Mathematically we want to search over all (noisy) mappings $g \rightarrow C$, holding $I(C; g)$ fixed and maximizing $I(C; x)$:
\begin{equation}
\max_{P(C|g)} \left[ I(C;x) - T I(C;g)\right],
\label{IB}
\end{equation}
where $T$ is a Lagrange multiplier that we introduce to fix $I(C; g)$.  This is the ``information bottleneck'' problem \cite{tishby+al_99}.  Buried inside the function that we are trying to maximize is  the joint distribution of positions and expression levels, $P(g, x)$, which has been estimated experimentally  \cite{petkova+al_19}, and it is this distribution that determines the form of the optimal $P(C|g)$.  

\begin{figure}
\centerline{\includegraphics[width = \linewidth]{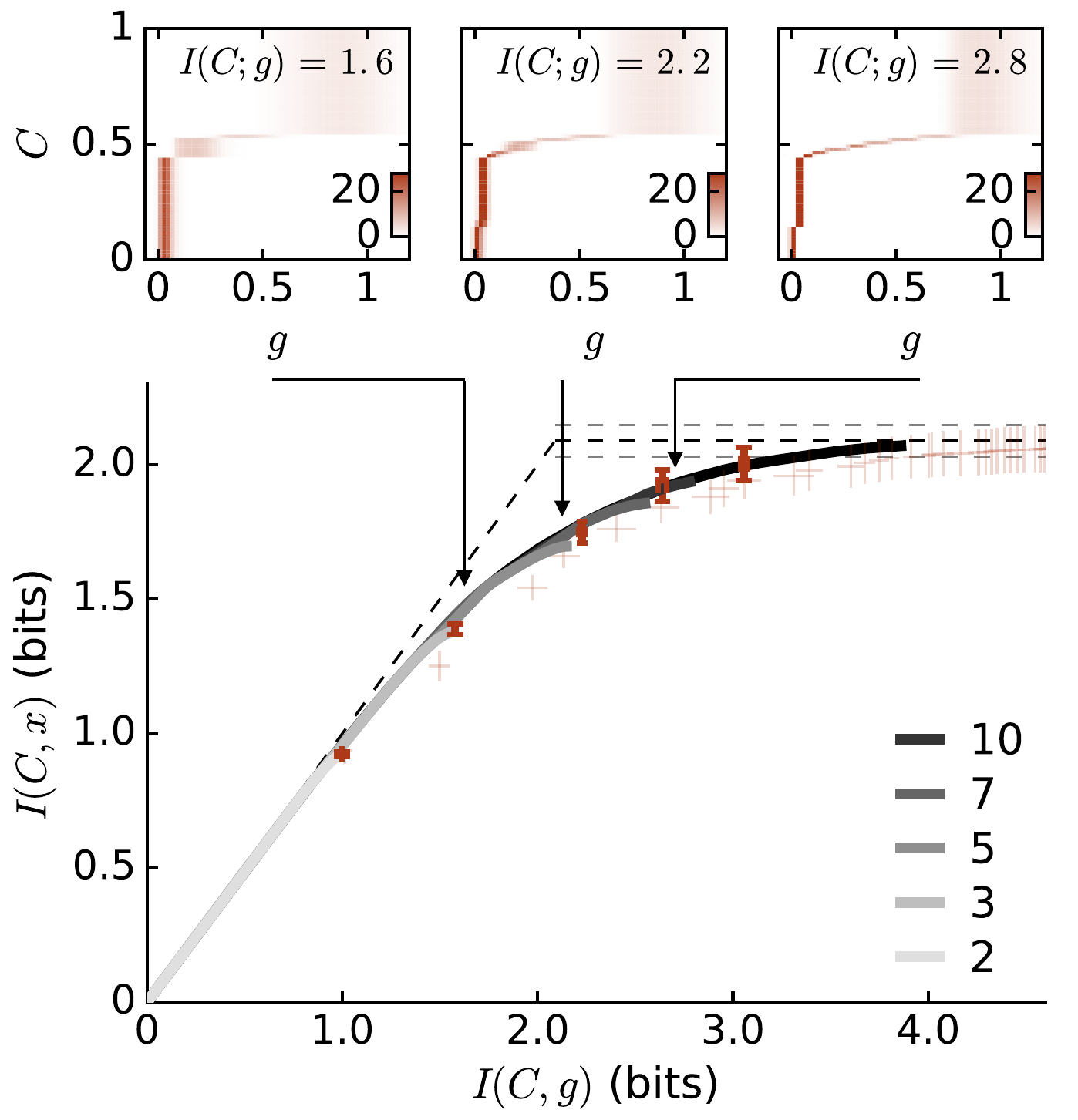}}
	\caption{The information bottleneck for positional information carried by Hb expression levels.  We map expression into some compressed description, $g \rightarrow C$, and find the maximum $I(C;x)$ at fixed $I(C;g)$, from Eq (\ref{IB}), shown as the solid line with different greyshades indicating different numbers of states $||C||$.  Solid points with error bars are the $T\rightarrow 0$ limit with fixed $||C||$, which reproduces the optimal discretization by multiple thresholds in Fig \ref{Fig01}E, and light points  are from an explicitly deterministic formulation of the bottleneck problem \cite{Slonim}. Upper panel shows snapshots  probability distributions $P(C|g)$ at different information capacities along the bottleneck curve; intermediate levels of $g \in [0.05,0.8]$ are progressively better resolved as the capacity increases.
\label{bottleneck}}
\end{figure}

We solve the optimization problem in Eq (\ref{IB}) numerically, considering $C$ to be a discrete set of states and varying the number of these states $||C||$.  At fixed $||C||$, decreasing $T$ allows $I(C;g)$ to be larger, and pushes the noisy mapping $P(C|g)$ toward being deterministic.  Results of the bottleneck analysis for Hb are shown in Fig \ref{bottleneck} as trajectories in the plane $I(C;x)$ vs $I(C;g)$.  The bounding curve that emerges from this analysis at large $||C||$ separates the plane into a physically possible region (below the curve) and an impossible region (above the curve).  As $I(C;g)$ becomes large, the curve plateaus at the available positional information $I(g;x)$, and must also stay below the diagonal, so that $I(C;x) \leq I(C;g)$.

The optimal thresholding models from Fig \ref{Fig01} correspond  to the endpoints of the bottleneck solutions with $||C||$ equal to the number of resolvable expression levels, and we see that these models  are almost on the optimal bounding curve.  This means that although the  picture of multiple noiseless thresholds is physically wrong, it does capture the impact of limited information capacity, almost quantitatively.  We can understand more about the structure of the optimal mappings $g\rightarrow C$ by looking at the distributions $P(C|g)$, shown in the top panels of Fig \ref{bottleneck}.  At small $I(C;g)$ the whole ranges of $g$ are mapped uniformly into ranges of $C$, while at larger $I(C;g)$ we see the emergence of a reliably graded mapping, especially in the range bracketing half--maximal expression.

We see  from Fig \ref{bottleneck} that capturing all the positional information encoded by Hb requires measuring the expression level with $\sim 3\,{\rm bits}$ of precision, consistent with our conclusions from the analysis of threshold models.  We have done the same analysis for the other gap genes (Kr, Gt, and Kni), with the same conclusion.
How does this compare with the information capacity of real genetic regulatory elements?  Estimates based both on direct measurements and on more detailed models indicate that this capacity is in the range of $1-3\,{\rm bits}$ \cite{tkacik+al_08a,tkacik+al_08b}.  This shows that  it is possible for cells to extract all the available positional information using a handful of enhancers, but this requires that they operate close to  the physical limits to information capacity.   At the same time,  it is clear from Fig \ref{bottleneck} that there is a very big difference between a capacity of $1\,{\rm bit}$ and $3\,{\rm bits}$.

The actual information capacity of genetic regulatory elements depends on the absolute concentrations of the relevant molecules, on the time available for reading out the information, and on other details of the different noise sources in the system \cite{tkacik+al_08a,tkacik+al_08b}.  At one extreme, if the capacity is three bits, then a single regulatory element is sufficient to capture the full positional information.  But if the capacity of a single element is only one bit, then we need multiple regulatory elements even in response to a single transcription factor.  

\begin{figure}[b]
\centerline{\includegraphics[width = \linewidth]{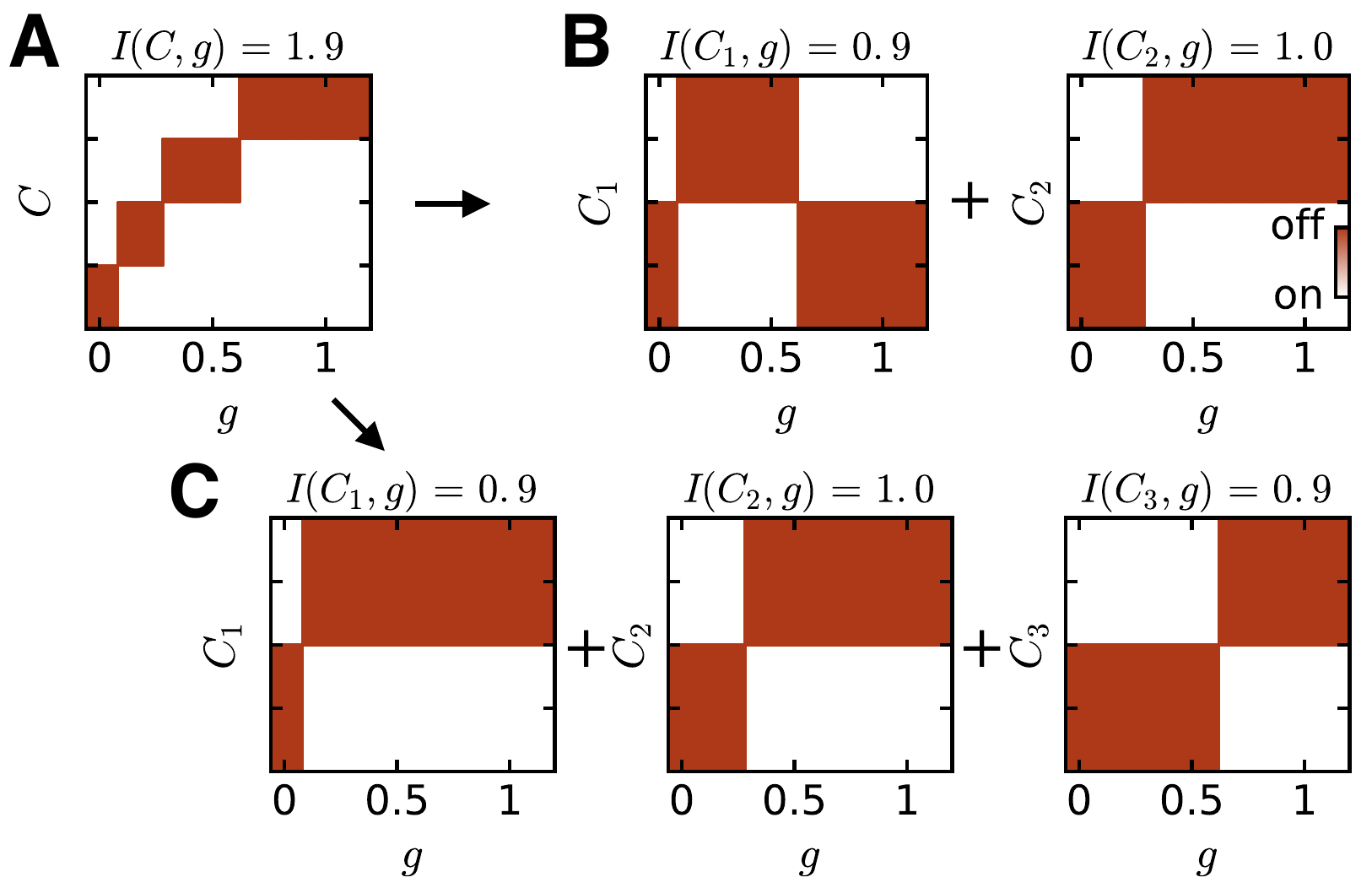}}
\caption{Decomposing a two bit element into one bit elements.  (A) Mapping from $g$ into four levels of $C$, capturing $I(C;g) = 1.9 \,{\rm bits}$.  (B) Decomposing $C = (C_1, C_2)$ where each $C_{\rm i}$ has two levels and captures $I(C_{\rm i}; g) \leq 1\,{\rm bit}$. This decomposition is exact, but the mapping $g\rightarrow C_1$ is non--monotonic.  (C) Decomposing $C = (C_1, C_2, C_3)$ where each $C_{\rm i}$ has two levels and captures $I(C_{\rm i}; g) \leq 1\,{\rm bit}$.  Adding a third element allows each mapping $g\rightarrow C_{\rm i}$ to be monotonic.
\label{sketch}}
\end{figure}

It might seem that if each element can capture $\sim 1\,{\rm bit}$ and the cell needs $\sim 3\,{\rm bits}$, then there must be three independent regulatory elements, but the problem is more subtle than this.  Three elements suffice if they provide independent information about the expression level $g$, but typically they carry overlapping or redundant information \cite{redundant}.    The only way to avoid this redundancy is for the mapping $g\rightarrow C$ be non--monotonic, as in  Fig~\ref{sketch}.  If we try to build a molecular implementation of non--monotonicity, we will end up with the transcription factor functioning (for example)  as an activator at low concentrations and a repressor at high concentrations.  There is older work suggesting that gap genes can act non--monotonically on their targets \cite{schulz+tautz_94}, and recent work motivated by bacterial systems showing how more complex patterns of TF binding and DNA looping can lead to non--monotonicity \cite{descheemaeker+al_19}.  There also is evidence that Hb in particular can act as an activator at some enhancer sites and a repressor at others \cite{staller+al_15}.

If the cell does not have access to molecular mechanisms that allow for non--montonicity, then there must be more of the $\sim 1\,{\rm bit}$ elements.  In the limit, we return to the picture of many elements, each of which switches at a different threshold, as discussed above.

\begin{figure}
\centerline{\includegraphics[width = \linewidth]{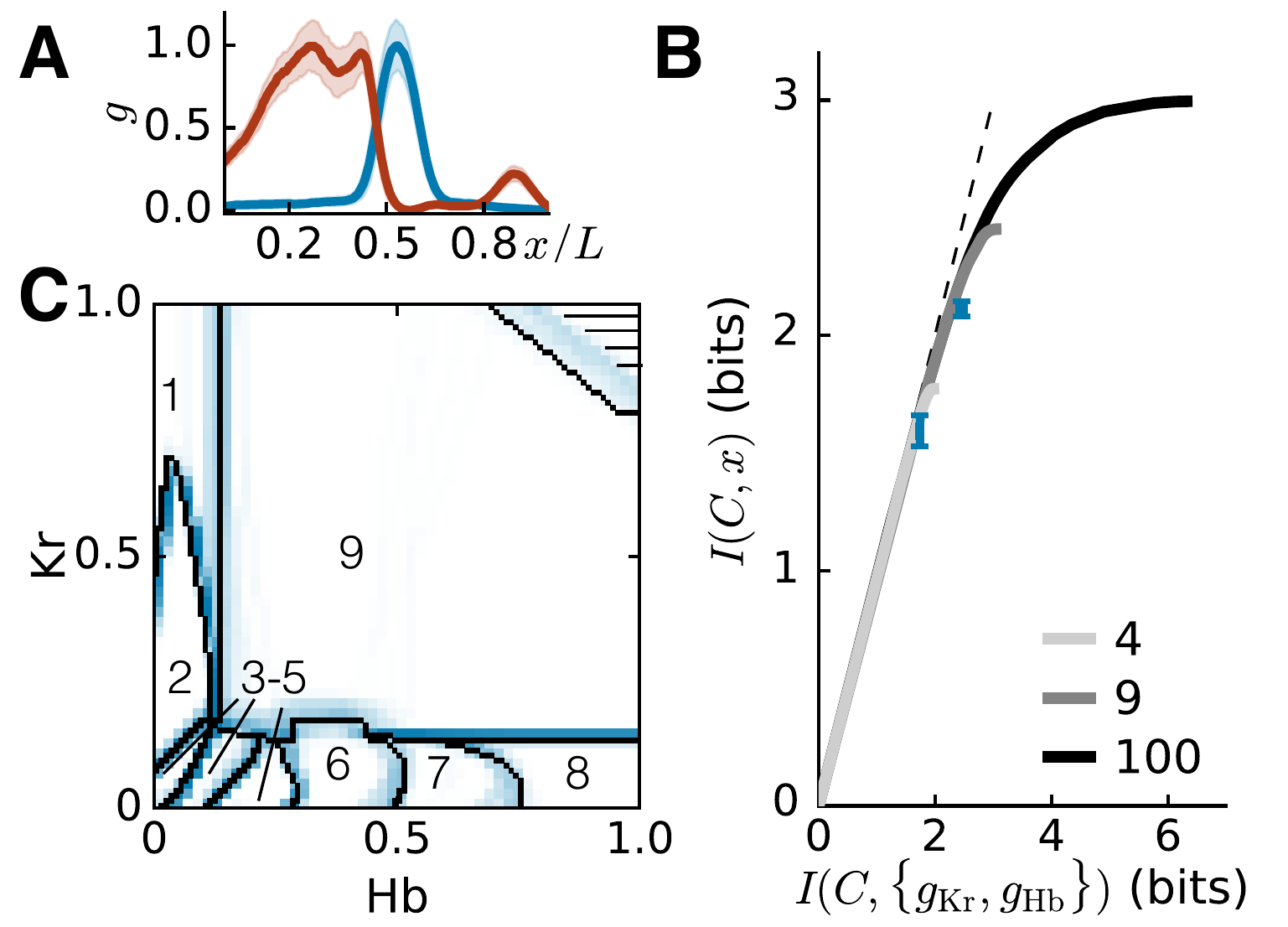}}
\caption{The information bottleneck for positional information carried by the combination of Hb and Kr expression levels. (A) Expression vs position along the anterior--posterior axis for Hb (red) and Kr (blue). Mean (solid) and standard deviation (shading) across $N_{\rm em} = 38$ embryos in a five minute window (40--44 min) in nuclear cycle 14 \cite{petkova+al_19}.  (B) Information bottleneck results, as in Fig \ref{bottleneck}, for $||C|| \in {4,9,100}$ (shades of grey), and optimal threshold solutions with one and two thresholds for Hb and Kr each (blue); thresholded solution are worse than the optimal bottleneck solutions.  (C) Structure of the optimal mapping $\{g_{\rm Hb} ,g_{\rm Kr}\} \rightarrow C$, with $||C|| = 9$ states in the limit $T\rightarrow 0$; solid black lines are boundaries between different states $C$,  blue shading indicates errors in our estimates of these boundaries, and hatched region at upper right is visited with nearly vanishing probability.
\label{HbKr}}
\end{figure}

Thus far our discussion has focused on the information carried by the expression level of just one gene, but we know that positional information is carried by the combined expression levels of all four gap genes \cite{dubuis+al_13,petkova+al_19}.   Rather than considering, as above, the mapping $g_{\rm Hb} \rightarrow C$, we can consider mappings from combinations of expression levels into compressed variables.  We start with Hb and Kr, so we consider  $\{g_{\rm Hb} ,g_{\rm Kr}\} \rightarrow C$.  The analog of Eq (\ref{IB}) is  the optimization problem
\begin{equation}
\max_{P(C|\{g_{\rm Hb} ,g_{\rm Kr}\})} \left[ I(C;x) - T I(C;\{g_{\rm Hb} ,g_{\rm Kr}\})\right].
\label{IB2}
\end{equation}
Aspects of the solution  are illustrated in Fig \ref{HbKr}.

The structure of the information bottleneck curve for two genes (Fig \ref{HbKr}B) is very similar to that for one gene.  To be sure that the positional information has saturated, we need to capture $\sim$6 bits of information about the combined expression levels of Hb and Kr, or $\sim 3$ bits per gene.  Importantly, the optimal compressions cannot decomposed into independent compressions of Hb and Kr.  With $\sim 6$ bits in total, the mapping $\{g_{\rm Hb} ,g_{\rm Kr}\} \rightarrow C$   divides the plane of expression levels into $||C|| \sim 2^6 = 64$ different regions, which is complicated to visualize.  For purposes of illustration we look at the optimal solution with $||C|| = 9$, shown in Fig \ref{HbKr}C.

The compressed state that covers the largest area in the $\{g_{\rm Hb} ,g_{\rm Kr}\}$ plane ($C =9$) has borders that are nearly parallel to the axes, so  this is an example of a state that could be defined by thresholding the two expression levels independently.  In contrast, some of the smaller areas have boundaries more nearly parallel to the diagonal, corresponding to thresholds on the ratio $g_{\rm Hb}/g_{\rm Kr}$.   These sorts of combinatorial structures become more prominent as we capture more of information, or consider more of the gap genes simultaneously.  In particular,  when we increase the cardinality (not shown), the largest area state (9) becomes sub--divided based on the ratio $g_{\rm Hb}/g_{\rm Kr}$, making it possible to capture the substantial amount of positional information that is available in the region where Hb and Kr expression levels cross, $x/L \sim 0.45$. This information is available only to mechanisms that can measure expression levels with high resolution, and the most efficient use of this resolution is to encode relative levels  rather than absolute levels.   This need for sensitivity to relative expression levels is the analog for multiple genes of the need for non--monotonicity in the case of a single gene.

In fact, mechanisms that are responsive to combinations of expression levels are quite natural.  Enhancer regions typically have closely spaced binding sites for multiple distinct transcription factors \cite{furlong+levine_18,StarkTrends}, and it is easy to imagine that binding events at these different sites interact.  Detailed models of the {\em eve} stripe enhancers show that activation reflects a competition between activation and repression by different gap gene proteins \cite{segal+al_08,crocker+al_16}, in qualitative agreement with a model in which the state of the enhancers serve as the compressed variable(s) $C$.

\begin{figure}
\centerline{\includegraphics[width = \linewidth]{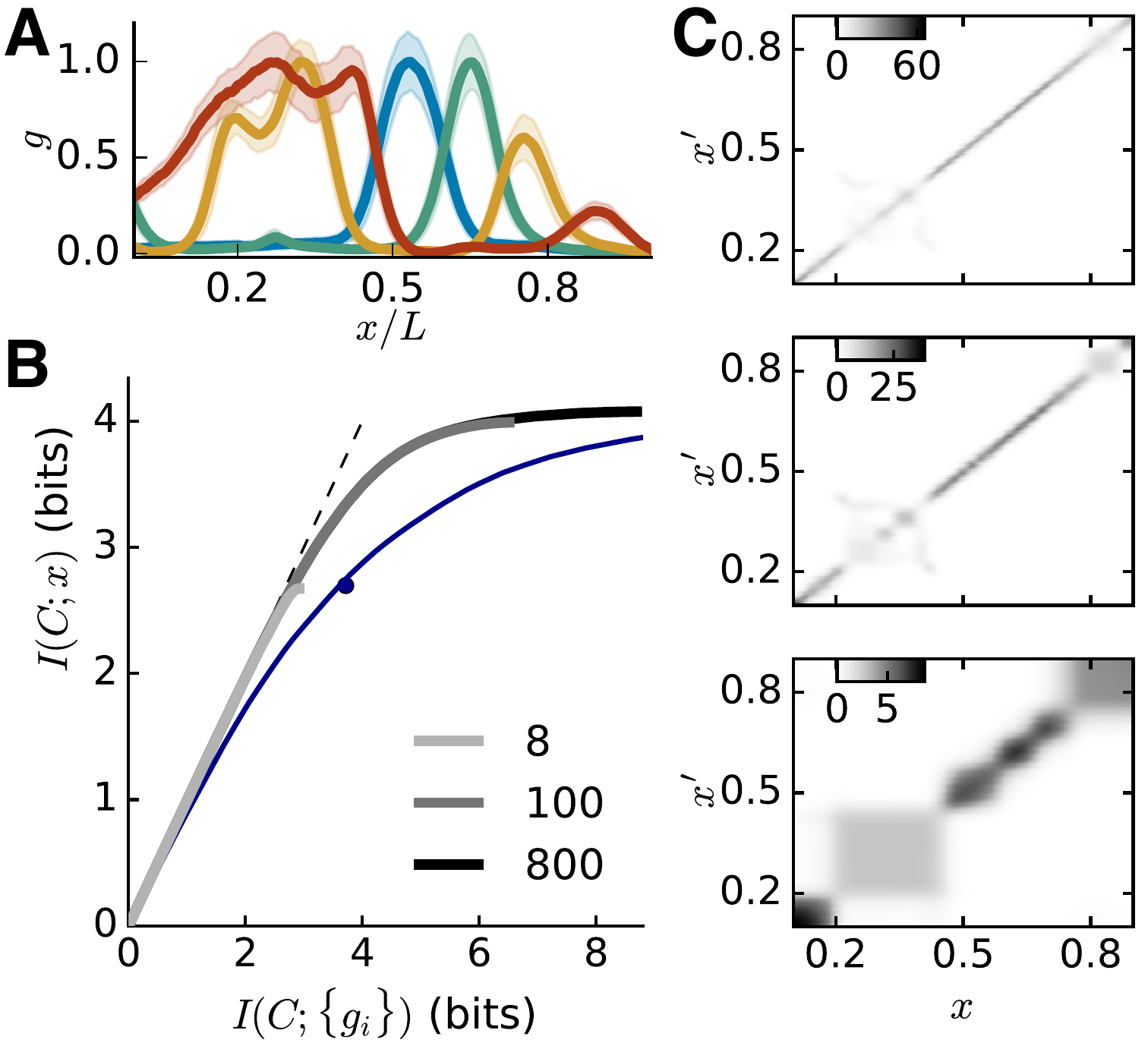}}
\caption{The information bottleneck for positional information carried by all four gap gene expression levels. (A) Expression vs position along the anterior--posterior axis for Hb (red), Kr (blue), Kni (green), and Gt (mustard).   Mean (solid) and standard deviation (shading) across $N_{\rm em} = 38$ embryos in a five minute window (40--44 min) in nuclear cycle 14 \cite{petkova+al_19}.  (B) Information bottleneck results, as in Fig \ref{bottleneck}.  Optimal solutions with $||C|| = 8,\, 100\,$ and $800$ (shades of grey), and solutions with independent compression of each gene expression level (blue); point corresponds to one but per gene.  (C) Decoding maps $P(x'|x)$  based on compressed representations of the expression levels. No compression  (top), $I(C; \{g_{\rm i}\}) = 4\,{\rm bits}$ (middle), and  $I(C; \{g_{\rm i}\}) = 2\,{\rm bits}$ (bottom).
\label{Fig4}}
\end{figure}

Finally we can ask what happens when we consider all four gap genes together, as in Fig \ref{Fig4}.  We know that in total there is more than four bits of positional information available, so that gap gene expression levels specify nuclear positions with $\sim$1\% accuracy along the length of the emrbyo \cite{dubuis+al_13}.    We can capture a large fraction of this information by keeping only four bits of information about the expression levels, or just one bit per gene (Fig \ref{Fig4}B).  But even this is not the same as reading expression level individually to one bit precision, since the optimal use of these four bits is to encode combinations of expressions levels, as in the case of two genes in Fig \ref{HbKr}.

To extract all the available positional information requires mechanisms that preserve eight or more bits of information about the combined expression levels of the four gap genes.  We can visualize what is being gained using the decoding maps introduced in Ref \cite{petkova+al_19}.  At the top of fig \ref{Fig4}C we consider cells at position $x$ that have access to the expression levels of all four gap genes, and use these data to infer their position $x'$; gray levels show the probability distributions $P(x'|x)$, which in this case form a narrow band around $x'=x$, with width $\sigma_x/L \sim 0.01$.  In the lower panels we imagine that inference is based not on the actual expression levels but on the compressed version $C$ that emerges from solving the bottleneck problem, and we do this for the optimal compressions with $I(C;\{g_{\rm i}\}) = 2,\,$ and $4\,{\rm bits}$.  We see that as the compression becomes more severe, the inference becomes more uncertain (larger $\sigma_x$) and genuinely ambiguous.

If we demand that compressed variables be constructed from individual gene expression levels, so that $C = \{C_{\rm i}\}$ and $g_{\rm Hb} \rightarrow C_1$, $g_{\rm Kr} \rightarrow C_2$, etc, then even with same total information capacity we capture less positional information, as indicated by the blue line in Fig \ref{Fig4}.  This loss of efficieny is substantial, and again indicates the importance of having regulatory mechanisms that are sensitive to combinations of expression levels.

To summarize, individual regulatory mechanisms  have limited information capacity \cite{tkacik+al_08b}, and our central result is that this capacity in turn sets strict limits on the amount of positional information that can be extracted from the gap gene expression levels.   Extracting all the available information requires that cells have an array of parallel responsive elements.  Further, to be efficient each element must respond not to individual gap genes signals but to combinations of these concentrations, so that the space of four concentrations is shattered into compact regions, generalizing the example of two genes shown in Fig \ref{HbKr}C.  In fact, the readout of positional information encoded in the gap genes is implemented by the array of enhancers controlling pair rule gene expression, and these enhancers are prototypical instances of regulatory elements that respond to combinations of transcription factors \cite{segal+al_08,crocker+al_16,furlong+levine_18}.  While there is some distance between our abstract formulation and the molecular details, it is attractive to see that this mechanistic complexity is required as a response to basic physical and information theoretic limitations.

\begin{acknowledgements}
We thank P-T Chen, M Levo, R Munshi, and B van Opheusden for helpful discussions.  This work was supported in part by the US National Science Foundation, through the Center for the Physics of Biological Function (PHY--1734030) and the Center for the Science of Information (CCF--0939370); by National Institutes of Health Grant R01GM097275; by the Alexander von Humboldt Stiftung; and by the Howard Hughes Medical Institute.
\end{acknowledgements}

\end{document}